\documentstyle[multicol,epsf,aps,prl,twoside]{revtex} % for two colums
\input epsf.tex

\begin{document}
\draft

\title{\ \\Aperiodicity-Induced Second-Order Phase Transition\\  
in the 8-State Potts Model}

\author{Pierre Emmanuel Berche, 
Christophe Chatelain,
 and Bertrand Berche\cite{byline1}}

\address{Laboratoire de Physique des Mat\'eriaux,~\cite{byline2} 
Universit\'e Henri Poincar\'e, Nancy
1, B.P. 239,\\ 
F-54506  Vand\oe uvre les Nancy Cedex, France} 

\date{November 18, 1997, to appear in Phys. Rev. Lett.}

\maketitle

\begin{abstract}
We investigate the  critical behavior of
the two-dimensional 8-state Potts model with an aperiodic distribution
of the exchange interactions between nearest-neighbor rows.  
The  model is studied numerically through intensive Monte Carlo 
simulations using the Swendsen-Wang cluster algorithm. The transition point is
 located through duality relations, and the critical behavior is
 investigated using FSS 
 techniques at criticality.
For strong enough fluctuations  of the aperiodic sequence under consideration, 
a second order 
phase transition is found. The  exponents $\beta/\nu$ and $\gamma /\nu$
 are obtained at the 
new fixed point. 
\end{abstract} 

%\pacs{PACS numbers: 64.60.Cn, 64.60.Fr, 05.50.+q, 05.70.Jk}

\markboth{{\rm Pierre Emmanuel BERCHE, Christophe CHATELAIN, and 
Bertrand BERCHE}\hfil 
\underline{{\rm \ }}}{\underline{{\rm \ }}\hfil {\rm Aperiodicity-Induced 
Second-Order 
Phase Transition\dots}} 

\begin{multicols}{2}
\narrowtext

The study of the influence of bond randomness on phase transitions is a quite
active field of research, motivated by  the importance of disorder in real 
experiments~\cite{cardy96c}.
According to the Harris criterion~\cite{harris74}, quenched randomness is a relevant
perturbation at a second order critical point when the specific heat exponent
$\alpha$ of the pure system is positive.

The analogous situation when the pure system exhibits a first order phase 
transition
was also studied.  Imry and Wortis first 
argued that quenched disorder could induce a second order phase 
transition~\cite{imrywortis79},
and it was  shown that in two dimensions,  an 
infinitesimal amount of randomly distributed
quenched impurities  changes the transition into a second order 
one~\cite{aizenmanwehr89-huiberker89}.
The first large-scale Monte Carlo study of  the effect of disorder 
at a   temperature-driven first order
phase transition is due to Chen, Ferrenberg, and 
Landau.
These authors studied the 2D 8-state  Potts model 
(which is known to exhibit a first order phase transition when the
number of states $q$ is larger than  
4~\cite{wu82}).
They first showed that the transition   becomes  
second order in the presence of bond
randomness, and   obtained   critical
exponents very close to those of the pure 2D Ising model 
at the new critical point~\cite{chenferrenberglandau9295}. 
On the other hand, drastically different results were obtained 
 for random lattices~\cite{jankevillanova96}.

The essential properties of random systems are governed by disorder 
fluctuations. All physical quantities depend on the configuration of disorder,
and the study of the influence of randomness requires an
average over disorder realisations. 
Among the systems where the presence of fluctuations is also of primary importance,
aperiodic systems have been of considerable interest since the
 discovery of quasicrystals~\cite{shechtman84}. They
are built in a deterministic way, making any configurational average
useless, and their critical 
properties have  been intensively studied (for a review, see 
Ref.~\onlinecite{grimm96}).
In layered systems, 
 aperiodic distributions of the exchange interactions between successive layers 
in the Ising model have been 
considered~\cite{igloi88-turban93}, leading to  unchanged universal behavior
or to modified critical properties, depending on the aperiodic series under
consideration.  The major result was obtained when 
Luck, generalizing  the Harris
criterion to layered perturbations, proposed a relevance criterion for the
fluctuating interactions~\cite{luck93a}.
According to Luck's criterion, 
aperiodic modulations may be relevant, marginal, or irrelevant, depending
on the correlation length exponent $\nu$ of the unperturbed system and on
a wandering exponent $\omega$ which characterizes the fluctuations of the couplings
around their average~\cite{queffelec87dumont90}. Systematic studies of the 
 critical properties for
irrelevant, marginal, and relevant aperiodic perturbations have then been achieved in the 
extreme anisotropic limit~\cite{turban94aetsuite}.

\begin{figure}
\epsfxsize=5.5cm
\begin{center}
\mbox{\epsfbox{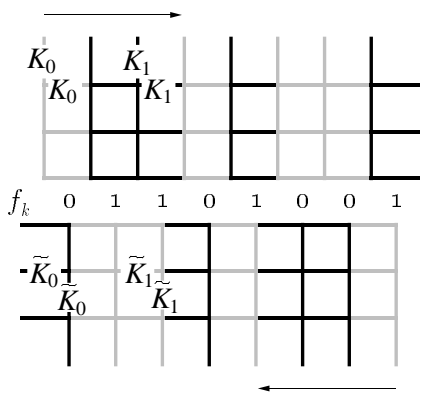}}
\end{center}
\caption{Layered aperiodic modulation of the coupling strengths on the
square lattice and dual system.}\label{couplages}  
\end{figure}

In this letter, we report  results of Monte Carlo simulations of
 the two-dimensional 8-state Potts 
model with an aperiodic
modulation of exchange couplings  between nearest-neighbor 
layers. Our aim is
to study the effect of such a  distribution on the nature of the phase
transition. In particular, we ask if the fluctuations
are able to induce a second order phase transition. 
The Hamiltonian of the system with aperiodic interactions can be written
\begin{equation}
-\beta {\cal H}=\sum_{(i,j)}K_{ij}\delta_{\sigma_i,\sigma_j} 
\label{eq-hamPotts}
\end{equation}
where the spins $\sigma_i$, located at sites $i$,  
can take the values $\sigma=1,2,\dots,q$ 
and the sum goes over nearest-neighbor pairs.
The coupling 
strengths are allowed to take two different values $K_0=K$ and $K_1=Kr$. They
are distributed according to a layered structure {\it i.e.} the distribution
is translation invariant in one lattice direction, and follows
an aperiodic modulation $\{ f_k\}$ of digits $f_k=0$ or $1$  in the other 
direction: In layer $k$, both horizontal and vertical couplings take the 
same value $Kr^{f_k}$ (Fig.~\ref{couplages}). The sequence of digits 
$\{ f_k\}$ is
 generated through iteration of 
substitution rules. 
The Thue-Morse (TM) sequence is obtained by 
  substitutions 
on digits:    $0\to {\cal S}(0)=01$, $1\to {\cal S}(1)=10$,
while the so-called paper folding sequence (PF) is generated through  
substitutions on pairs of digits: $11\to {\cal S}(11)=1101$, 
$10\to {\cal S}(10)=1100$, $01\to {\cal S}(01)=1001$,
$00\to {\cal S}(00)=1000$. 
After 3 iterations initiated by 0 and 11 respectively, we
get  the following sets $\{ f_k\}$
\begin{equation}
 {\rm TM}: 01101001,
\quad {\rm PF}:1101100111001001
\label{eq-PF}
\end{equation}
Most of the properties of a sequence are  obtained from the substitution
matrix~\cite{queffelec87dumont90}.
The asymptotic density $\rho_\infty$ of 1,  
the 
length $L_n$ of the sequence after $n$ iterations, 
but also the fluctuations of the $f_k$'s  
at a length scale $L_n$  around their average values 
are related to the substitution 
matrix. For the fluctuations, one has
 \begin{equation}
\sum_{k=1}^{L_n}(f_k-\rho_\infty)\sim L_n^\omega
\label{eq-fluct}
\end{equation}
where  $\omega$, 
 the wandering exponent, discriminates between bounded and unbounded
 fluctuations. 
In the case of TM and PF sequences, the fluctuations are 
respectively non-divergent ($\omega_{{\rm TM}}=-\infty$), and 
logarithmically divergent
($\omega_{{\rm PF}}=0$).

Our particular choice of coupling distribution makes it possible to determine
exactly the critical point by duality arguments. Consider a system of $L$ layers with a
distribution $\{f_k\}$, made from a succession of vertical-horizontal (V-H) 
bonds
when read from left to right (Fig.~\ref{couplages}), and let us write its 
singular free energy density
$f_s(K_0,K_1;\{f_k\})$.
Under a duality transformation, the strong and weak couplings $K_i$ are 
replaced by 
weak and strong dual couplings $\tilde K_i$ respectively. Since a vertical bond on 
the original
lattice becomes horizontal on the dual system, the same V-H bond configuration 
is 
recovered for the transformed system when the distribution is read from 
right to left.
 One thus
gets the same type of system, but a reverse distribution 
$\{f_{L+1-k}\}$, so that
the free energies of the two systems are the same: $f_s(K_0,K_1;\{f_k\})=
f_s(\tilde K_0,\tilde K_1;\{f_{L+1-k} \})$.
The sequences considered here have the interesting property that the reverse
 distribution
corresponds to the original one if one interchanges perturbed and unperturbed
couplings $K_1\leftrightarrow K_0$: $f_s(\tilde K_0,\tilde K_1;\{f_{L+1-k} \})=
f_s(\tilde K_1,\tilde K_0;\{f_k\})$.
The system being self-dual
the critical point, if  unique,  is exactly given by the critical 
line $K_{0c}=\tilde K_{1c}$
of the usual anisotropic model~\cite{fisch78-kinzeldomany81}:
\begin{equation}
({\rm e}^{K_c}-1)({\rm e}^{K_cr}-1)=q. 
\label{eq-Kc}
\end{equation}
One should mention there that the required symmetry property of the sequences
holds for odd iterations in the case of TM, and works in the case of PF if one 
omits
the last digit, which simply introduces an irrelevant surface effect in the
simulations.

We performed extensive simulations of $L\times 2L$ lattices 
($16\leq 2L\leq 512$)
with periodic boundary conditions in one direction (vertical) and 
free boundaries in the 
other ($2L$ columns). The  Swendsen-Wang cluster 
flipping method~\cite{swendsenwang87} 
was used. Between $2\times10^5$ (smaller  lattice sizes)
 and $6\times10^5$ (larger  lattice sizes) Monte Carlo steps (MCS) per spin were 
 performed (this is always larger than $10^4$ times 
 the correlation time, and seems sufficient in order to produce reliable
 thermal averages).

The  order parameter is defined by the majority orientation of the 
spins~\cite{challalandaubinder86}:
\begin{equation}
M=\langle m\rangle=\frac{q\rho_{max}-1}{q-1}.
\label{eq-locordpara}
\end{equation}
Here, $\rho_{max}=\langle max_\sigma(\rho_\sigma)\rangle$, where $\rho_\sigma$ 
is the density of spins in the state $\sigma$ and $\langle \dots
\rangle$ denotes
the thermal 
average over the Monte Carlo iterations. The susceptibility is given by $\chi=
KV (\langle m ^2\rangle-\langle m \rangle^2)$.
Although local ordering mechanisms are not yet  clarified  in
 aperiodic systems, we expect a unique
transition temperature for
all the columns, so we used average quantities
 in order to reduce
fluctuations.

\begin{figure}
\epsfxsize=8.5cm
\begin{center}
\mbox{\epsfbox{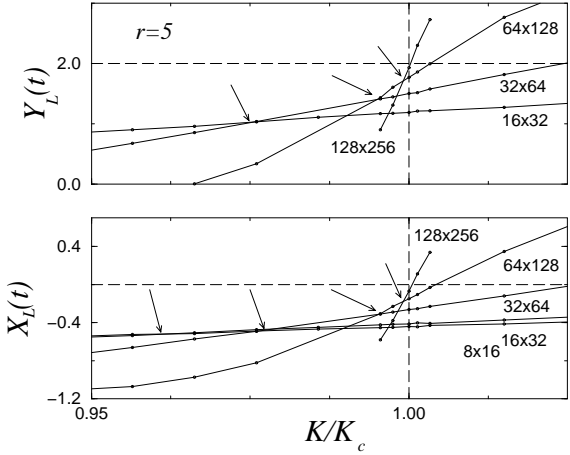}}
\end{center}
\vskip -5mm\caption{Temperature-dependent effective exponents for 
the average magnetization
and  suceptibility (TM sequence) estimated from the ratios ${X}_L(t)=\frac{\ln M_L/M_{L/2}}{\ln 2}$
and ${Y}_L(t)=\frac{\ln \chi_L/\chi_{L/2}}{\ln 2}$ ($K_1/K_0=r=5$). The sizes 
vary from $2L=16$ to $2L=256$, and the arrows indicate the crossings between
successive sizes which give rough estimates for the critical exponents.
}\label{effectiveTM}  
\end{figure}

The first task is to identify the order of the transition. For that purpose,
we made some preliminary runs at several temperatures first of all in order
to confirm numerically the location of the critical point, and then to have
a general picture of the phase transition.
The examination of the energy autocorrelation time shows
 that it is
diverging in the case of TM sequence, while it seems to remain more or less 
bounded for PF. It is consistent with a first order transition in the first 
case, and
a second order one for the latter. We have further   estimated temperature-dependent effective exponents for 
the average magnetization
and  suceptibility. This can be done by comparing the data at  two different
sizes $L$ and $L'=L/2$: Assuming the following scaling form for the average
magnetization 
 $M_L(t)=L^{-\beta/\nu}{\cal M}(Lt^\nu)$,
 where $t=\mid\! K-K_c\!\mid$, we define the quantity~\cite{henkelherrmann90}
 \begin{equation}
{X}_L(t)=\frac{\ln M_L/M_{L'}}{\ln L/L'}.
 \label{eq-X}
 \end{equation}
 Close to $K_c$, this can be 
 expanded in powers of $Lt^\nu$, leading to 
 \begin{equation}
 {X}_L(t)\simeq -\frac{\beta}{\nu}+\frac{Lt^\nu}{2\ln 2}
 \frac{{\cal M}'(Lt^\nu)}{{\cal M}(Lt^\nu)}+O(L^2t^{2\nu})
 \label{eq-effective}
 \end{equation}
which defines an effective exponent. As the critical point is approached and in the 
thermodynamic
 limit, it evolves towards $-\frac{\beta}{\nu}$. 
 The analogous quantity can be computed for the 
 susceptibility. The results are shown for TM on Fig.~\ref{effectiveTM}.

\vskip -5mm

 The successive estimates of  $\frac{\beta}{\nu}=d-y_t$ and 
 $\frac{\gamma}{\nu}=2y_h-d$
 clearly  evolve 
 towards the values
 0 and 2, characteristic of a first order phase transition. The scaling
dimensions associated to the temperature and magnetic field, $y_t$ and $y_h$,
indeed take a special value equal to the dimension $d$ of the 
system~\cite{fisherberker82}.
  In the case of the PF sequence,
 the behavior is drastically different, and this first analysis  does not
 allow any conclusion.

\begin{figure}
\epsfxsize=8.5cm
\begin{center}
\mbox{\epsfbox{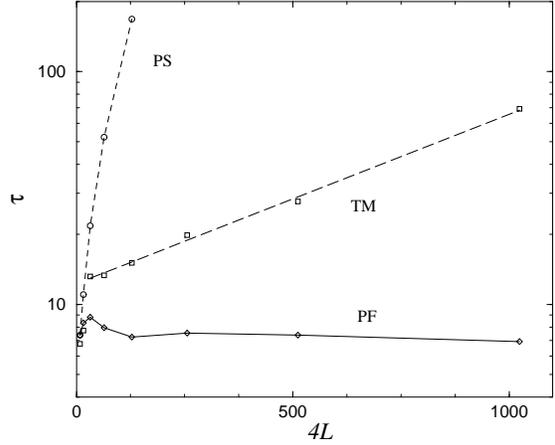}}
\end{center}
\vskip -5mm\caption{Energy autocorrelation time
$\tau$ at $K^c$ ($r=5$). For TM, the dashed line is a fit to an exponential behavior.}
\label{autoc}  
\end{figure}

 Once the qualitative description of the phase transition was made, our strategy was to
use finite-size scaling (FSS) techniques in order to get more accurate results. We made runs
for systems of larger sizes, and in a $L\times 4L$ geometry ($4L$ in the aperiodic direction,
going from 8 to 1024),
for which we estimated the number of MCS/spin from the preliminary runs. We have moreover
studied a periodic system (PS) with alternate couplings $K_0$ and $K_1$ 
(i.e. the same critical
point given in Eq.~\ref{eq-Kc}), and which is 
a ``first order reference'' system.
 This is illustrated in Fig.~\ref{autoc} where the energy autocorrelation time
$\tau$ is plotted for the three samples. In the case of TM, $\tau$
diverges exponentially as expected for a first order phase transition, 
although it is
always quite small compared to the periodic system. For PF, the data are
compatible with a power law with a very small dynamical exponent as expected
 for cluster algorithm simulations at a second order 
phase transition.
   
\vskip -6mm  
\vskip -1.2cm 

\begin{figure}
\epsfxsize=8.5cm
\begin{center}
\mbox{\epsfbox{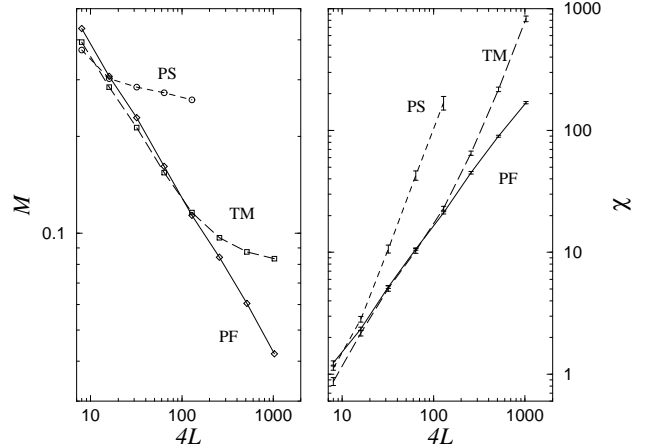}}
\end{center}
\vskip -5mm\caption{Log-log plots of $M$ and $\chi$ vs $L$ for a periodic 
reference, and
for TM and PF sequences ($r=5$). For $M$, error 
bars are smaller than the symbol sizes.}
\label{FSS}  
\end{figure}
\vskip -1mm

\begin{figure}
\epsfxsize=8.5cm
\begin{center}
\mbox{\epsfbox{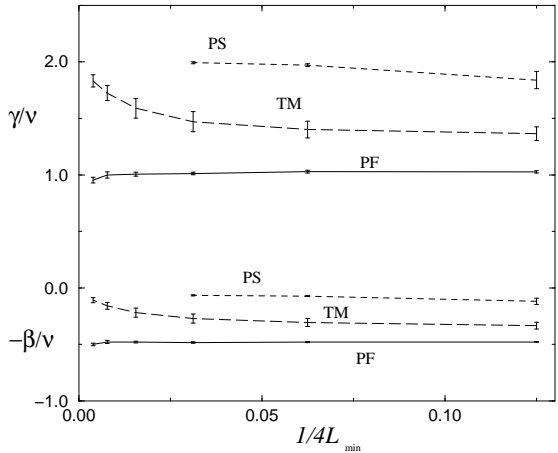}}
\end{center}
\vskip -5mm\caption{Effective size-dependent exponents for the two 
aperiodic sequences ($r=5$).}
\label{eff-L}  
\end{figure}

 The crude  data for $M$ and $\chi$ furthermore show that, in the case 
 of the TM sequence,
a cross-over appears between small sizes where the data more or less follow
the same behavior than PF, and large sizes where the first order regime
analogous to PS is 
well established (Fig.~\ref{FSS}).
Certainly,  a careful procedure is needed for a reliable determination
 of the critical exponents.
 From the log-log curves between $4L_{min}$ and $4L_{max}=1024$, one
  determines 
an effective
exponent $x(L_{min})$ for each quantity, 
then the smaller size is cancelled from the data 
and the whole procedure is 
repeated until the three largest sizes only remain. The effective exponent is 
then
plotted against $L^{-1}$ (Fig.~\ref{eff-L}). The critical exponent is finally
deduced
from the  extrapolation at infinite size. 
The numerical results are given in Table~\ref{tab1}. All of them are in 
agreement with the 
scaling law $d=2\beta/\nu +\gamma/\nu$, within the precision of the results.

To summarize, we have shown from numerical simulations that the fluctuations
introduced by an   aperiodic modulation  of exchange 
interactions is liable to
induce a second order phase transition in a system which
originally exhibits a first order transition. 

\vskip -5mm 

\vskip -5mm
\vbox{
\narrowtext
\begin{table}
\caption{Exponents associated to the magnetization and susceptibility for the 
three examples considered in the text. The  uncertainties given there are rough estimations corresponding to the 
standard
deviation 
for the fit of the data in the whole range of sizes.
\label{tab1}}
\begin{tabular}{lccc}
  & PS & TM & PF \\
\tableline
$ {\beta}/{\nu}$ & $0.05\pm 0.03$ & $0.05\pm 0.04$ & $0.48\pm 0.03$ \\
$ {\gamma}/{\nu}$ & $1.99\pm 0.08$ & $1.96\pm 0.06$ & $1.01\pm 0.04$ \\
\end{tabular}
\end{table}
\narrowtext
 
 }\vskip -5mm

From Monte Carlo simulations, we  have strong evidences in favor
of a second order regime for PF sequence.  This type of effect was already known
since the work of Imry and Wortis in the case of a random distribution where the
fluctuations are unbounded and can be characterized by a wandering exponent
$\omega_{{\rm rand}}=1/2$. Here, the same type of behavior is induced
by a smoother perturbation, namely
by the PF   sequence which exhibits only logarithmic fluctuations ($\omega=0$)  
while the bounded 
fluctuations  generated by the TM sequence ($\omega=-\infty$) are not strong 
enough
 to destroy the first order transition.
 The same type of problem  on a 
 quasicrystal
 is currently under investigation by Ledue et al and the transition seems, in this 
 case also,
  to 
 remain of
 first order~\cite{ledue97}.
We may thus infer that Luck's criterion can probably be applied to 
first order phase
transition. 
 Here, we can replace $\nu$ by $1/d$~\cite{fisherberker82,janke92} in 
 the criterion in order 
 to compare thermal fluctuations
 to those introduced by the distribution of couplings. Luck's 
 cross-over exponent
 then becomes $\phi=1+(\omega -1)/d$, and the aperiodicity can 
 induce a second order
 phase transition when $\phi >0$. This is in 
 agreement with the results of our simulations. 
We can finally mention that the study of local order parameter is currently
under investigation.

We  are indebted to L. Turban, W. Janke, and D. Ledue
 for valuable discussions.  This work was supported by the Ciril and the Centre 
 Charles Hermite in Nancy.

\end{multicols}

\end{document}